\documentclass[a4paper,11pt]{article}
\usepackage{pos}
\usepackage[section]{placeins}

\title{Agent-based code generation for the Gammapy framework}

\author*[a]{Dmitriy~Kostunin}
\author[b]{Vladimir~Sotnikov}
\author[c]{Sergo~Golovachev}
\author[a]{Abhay~Mehta}
\author[a]{Tim~Lukas~Holch}
\author[a]{Elisa~Jones}

\affiliation[a]{Deutsches Elektronen-Synchrotron DESY, 15738 Zeuthen, Germany}
\affiliation[b]{JetBrains Limited, 8046 Paphos, Cyprus}
\affiliation[c]{JetBrains GmbH, 80639 München, Germany}

\emailAdd{dmitriy.kostunin@desy.de}

\newcommand{\infoicon}[1][1.2em]{%
  \raisebox{-0.3\height}{\includegraphics[height=#1]{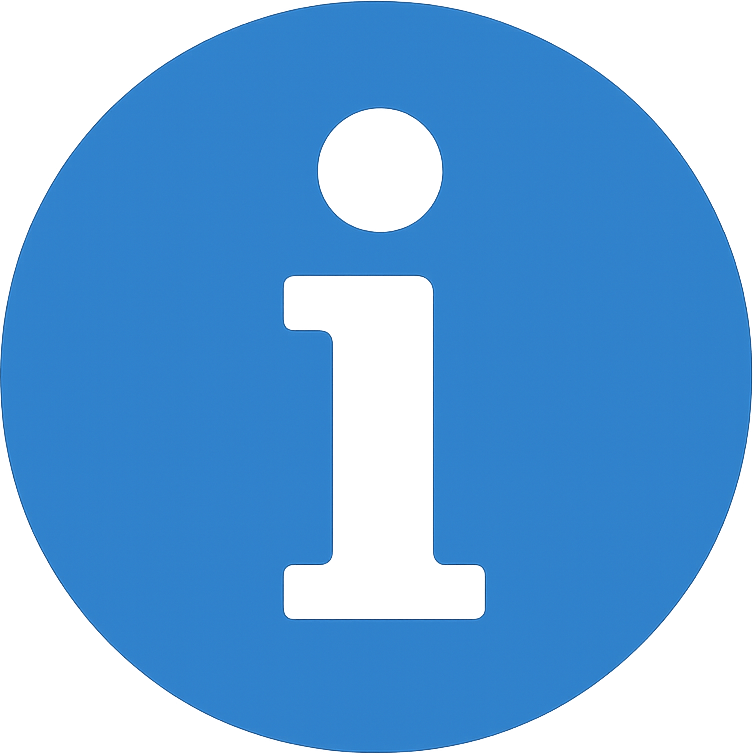}}%
}
\abstract{
\infoicon~The proceedings text was generated with a large language model (LLM), using the agent’s source code as a reference.
~\\
~\\
Software code generation using Large Language Models (LLMs) is one of the most successful applications of modern artificial intelligence. Foundational models are very effective for popular frameworks that benefit from documentation, examples, and strong community support. In contrast, specialized scientific libraries often lack these resources and may expose unstable APIs under active development, making it difficult for models trained on limited or outdated data. We address these issues for the Gammapy library by developing an agent capable of writing, executing, and validating code in a controlled environment. We present a minimal web demo and an accompanying benchmarking suite. This contribution summarizes the design, reports our current status, and outlines next steps.
}

\ConferenceLogo{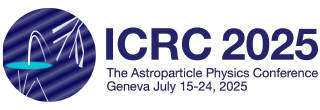}
\FullConference{39$^{\mathrm{th}}$ International Cosmic Ray Conference (ICRC2025)\\
15--24 July 2025, Geneva, Switzerland}

\begin{document}
\maketitle


\section{Introduction \label{sec:intro}}
Large Language Models (LLMs)~\cite{zhao2025surveylargelanguagemodels} are increasingly capable of writing non-trivial scientific software, but turning free-form assistance into reproducible \emph{analysis scripts} still requires domain knowledge and guard-rails~\cite{2025arXiv250321460L, 2025arXiv250814111W}\footnote{\url{https://github.com/luo-junyu/Awesome-Agent-Papers}}\footnote{\url{https://agenticscience.github.io/}}. In the domain of gamma-ray astronomy this gap is amplified by: (i)~domain-specific data formats (Data Level~3, DL3), (ii)~fast-moving APIs, and (iii)~the need to run code in a controlled environment with the correct data mounted~\cite{2025arXiv250300821K}.

\textbf{Gammapy}~\cite{gammapy:2023} is an open-source Python package for high-energy gamma-ray astronomy built on the scientific Python stack. It provides a uniform platform to reduce and model DL3 data from different instruments, with background-estimation methods and Poisson maximum-likelihood fitting, and is used across the community, including next-generation Cherenkov Telescope Array Observatory (CTAO) tooling.

We present an \textbf{agent for Gammapy} that generates, executes, and self-repairs analysis scripts until they run successfully. The agent follows three principles:

\begin{itemize}
  \item \textbf{Strong contracts in prompting.} The system message encodes rules such as “return one complete Python script”, “import all dependencies”, “do not call plotting display functions”, and “do not select observations via \texttt{TARGET\_NAME}”. Rules are enforced at generation time and by validation.
  \item \textbf{Tight integration with the analysis stack.} The agent exposes the data location via an environment variable and executes code in a sandboxed workspace; stdout/stderr are captured, and timeouts protect against runaway jobs.
  \item \textbf{Iterative verification.} Generated code is executed; failures are summarized and fed back to the model. The loop stops when validation succeeds or a configured attempt budget is exhausted.
\end{itemize}

The implementation is a small, testable Python package (\texttt{gammapygpt}) with a benchmarking suite and a minimal web user interface (UI). It targets everyday Gammapy tasks (listing observations, spectral extraction with reflected regions, 3D binned analyses, quick-look maps), prioritizing clarity over cleverness. The text below reflects the not yet public code accompanying this paper.

\section{Agent design for Gammapy analysis \label{sec:design}}
\begin{figure}[t]
  \centering
  \includegraphics[width=0.48\textwidth]{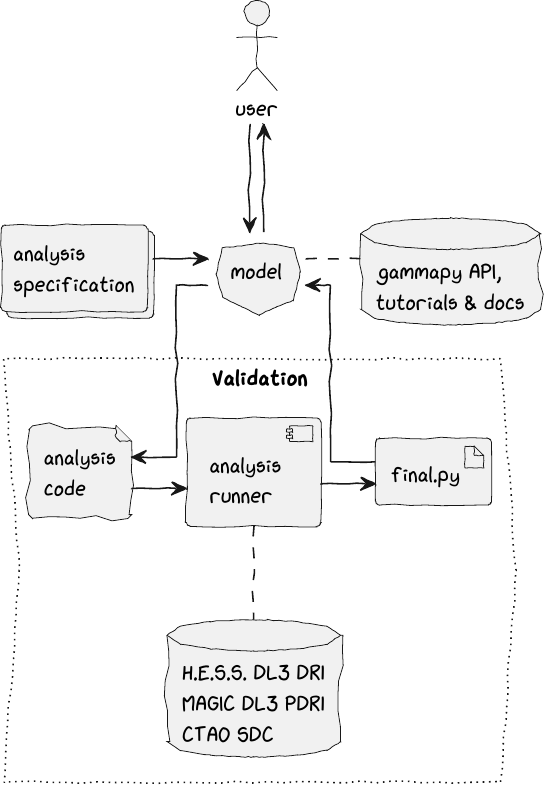}~~
  \includegraphics[width=0.48\textwidth]{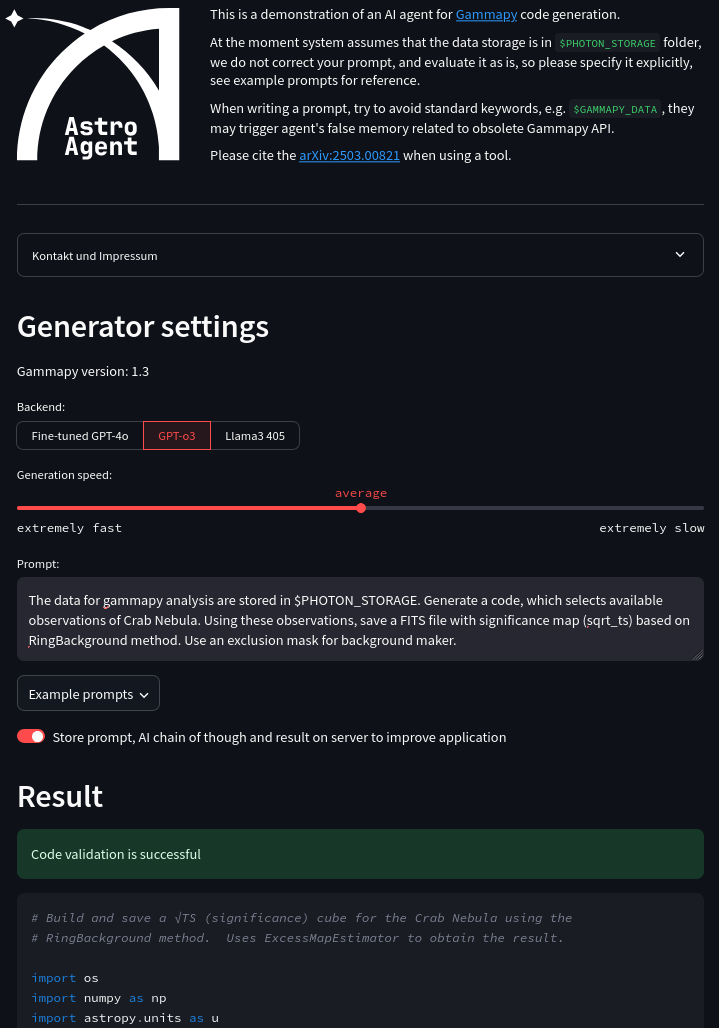}
  \caption{\textit{Left:} Block diagram of the agent. Solid arrows form the generation–execution–validation loop; dashed arrows indicate retrieval of contextual snippets (tutorials, examples). \textit{Right:} Screenshot of the Streamlit prototype (\url{https://majestix-vm8.zeuthen.desy.de}).}
  \label{fig:overview}
\end{figure}

Figure~\ref{fig:overview} sketches the workflow. A user prompt is expanded into a governed chat history; the model returns a single Python script; the script is executed in a controlled environment; the result is validated and, if needed, a concise error summary triggers the next iteration.

\subsection{System architecture}\label{sec:arch}
The repository is organized in small modules:

\paragraph{Configuration.}
A \texttt{Config} model stores \emph{backends} (model name, base URL, API key, embedding model, and the preferred reasoning effort), timeouts, and filesystem locations. The data location is published via the environment variable \texttt{PHOTON\_STORAGE}; users can override paths at runtime. A simple CLI saves/loads configuration files and provides helpers to fetch public H.E.S.S.\ DL3 data~\cite{HESS:2018zix}.

\paragraph{Messages and prompting.}
The system message encodes non-negotiable rules:
\begin{itemize}
  \item return \emph{exactly one} Python code block (no prose),
  \item import all requirements; avoid interactive plotting,
  \item avoid selecting observations via \texttt{TARGET\_NAME},
  \item prefer current Gammapy idioms.
\end{itemize}
For compatibility with tutorials and examples the system text still mentions \texttt{GAMMAPY\_DATA}, while the agent itself publishes \texttt{PHOTON\_STORAGE}. Both are accepted by the validator.

\paragraph{Runner and tool use.}
The \emph{Runner} constructs the conversation (system + user + optional RAG context), calls the model with the configured reasoning effort, and enforces a lightweight “\emph{return\_python}” tool so the output is a single script without backticks or prose. The script is executed and validated. On failure, the Runner appends a new user turn containing a compact traceback tail and hints; iteration continues until success or the attempt budget is reached.

\paragraph{Code execution.}
Scripts are written to a temporary file and executed under a pared-down environment that includes the data pointer (\texttt{PHOTON\_STORAGE}). Stdout/stderr are captured, and a hard time limit aborts runaway jobs. When a \texttt{prefix} directory is configured, each attempt persists the generated script and the chat transcript for auditability.

\paragraph{Retrieval-augmented generation.}
An optional RAG layer indexes selected Gammapy tutorials into an in-memory Qdrant\footnote{\url{https://qdrant.tech/}} collection using OpenAI embeddings. Top-$k$ snippets (subject to a score threshold) are injected as additional user context. The index can be rebuilt deterministically from the same sources.

\paragraph{Utilities and data handling.}
Notebook-oriented tutorial scripts are pre-processed (plots stripped; IPython magics removed) to respect the “no plotting” contract and ensure headless execution. All data access in generated code is expected to resolve via \texttt{PHOTON\_STORAGE}.

\subsection{Interfaces}
Two thin interfaces sit on top of the same core:
\begin{itemize}
  \item a \textbf{CLI} with sub-commands to generate code, download datasets, and (optionally) build the RAG index;
  \item a minimal \textbf{Streamlit}\footnote{\url{https://streamlit.io/}} web app where users can choose a backend, set the generation attempt budget via a “speed” slider, toggle message persistence, and run prompts. The app displays the latest script and execution log and stores the conversation in a run folder.
\end{itemize}

\subsection{Validation and safety}
Validation is pluggable. By default an execution is \emph{valid} if the process returns code~0; additional domain checks are supported (e.g.\ verifying expected numerical outputs). All runs are \emph{offline}: the executor has no network access and only sees the mounted data path. Timeouts and filesystem prefixes are configured centrally, making runs reproducible.

\subsection{Reproducibility}
Each attempt yields a folder containing: the prompt, the message log, the generated script, raw stdout/stderr, and the validation outcome. This turns successful generations into reusable examples and keeps failures easy to diagnose.

\section{Benchmarks}
Benchmarking is essential to quantify model performance for \texttt{Gammapy} code generation. We include a harness in \texttt{gammapygpt.benchmark} that executes generated scripts in isolated environments, records iteration traces, and applies numerical validators with explicit tolerances where needed. We compare two OpenAI reasoning models, \texttt{o3} and \texttt{gpt-5}~\cite{openai2025gpt5systemcard}, both at the highest available reasoning effort.

\begin{figure}[b]
    \centering
    \includegraphics[width=1\linewidth]{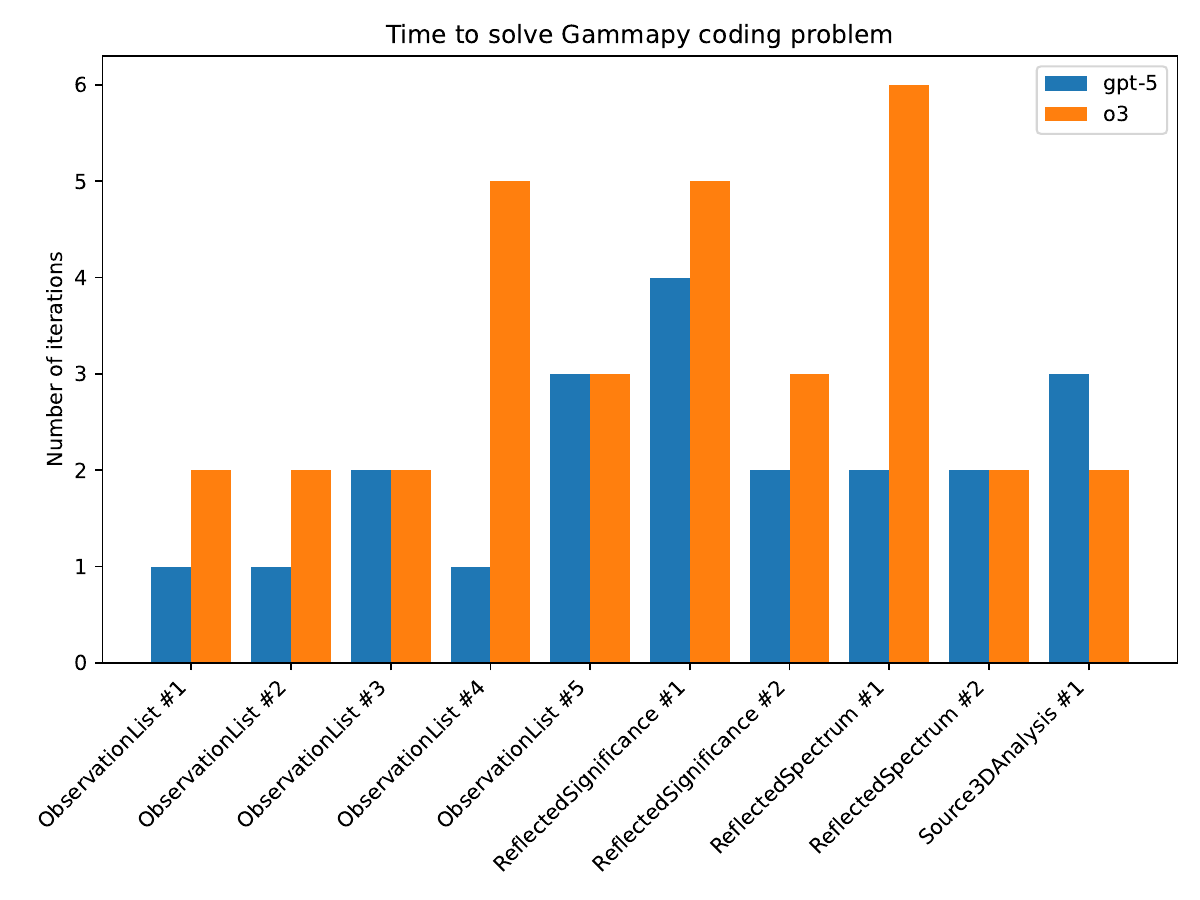}
    \caption{Coding benchmark results (attempts to pass and pass rates per task/model).}
    \label{fig:benchmarks}
\end{figure}

\paragraph{Tasks and criteria.}
Each task consists of (i) producing runnable code and (ii) meeting a task-specific check. Examples include:
\begin{itemize}
  \item \textit{ObservationList}: select observations for a given source and print the number of observations (exact integer match).
  \item \textit{ReflectedSignificance}: compute a reflected-region significance for a source.
  \item \textit{ReflectedSpectrum}: perform a reflected-region spectral extraction and report the total energy flux and spectral index (floats with tolerance).
  \item \textit{Source3DAnalysis}: reduce all available observations of a source to a \texttt{MapDataset} and fit a spatial–spectral model (considered a pass if the script runs end-to-end under the timeout, due to complexity).
\end{itemize}

\paragraph{Harness details.}
Every iteration logs exception classes, a compact traceback tail, tokens (input, cached input, output, and reasoning tokens), and the attempt index at which validation passed. RAG is optionally enabled, injecting top-$k$ tutorial snippets when available.

\paragraph{Results.}
On the smaller \emph{per-source} tasks both models with high reasoning effort reached 100\% pass rate in our runs, the most recent model shows slightly faster performance.
The histogram in Figure~\ref{fig:benchmarks} summarizes attempts-to-pass across tasks. As an illustration of trace logging, a typical successful run recorded 7.3k output tokens of which \(\sim\)6.5k were reasoning tokens.

\section{Conclusion \label{sec:conclusion}}
Domain-aware code agents are already practical for gamma-ray astronomy. By aligning a state-of-the-art reasoning model with up-to-date \texttt{Gammapy} workflows and wrapping it in a strict validation loop, our agent delivers reliable analysis scripts and reduces boilerplate for routine tasks.

In parallel, we are extending the backend layer to support \emph{open-weight} models for on-premise and privacy-preserving deployments. Concretely, we are operating and evaluating Qwen~\cite{2025arXiv250509388Y} and OpenAI’s GPT-OSS~\cite{2025arXiv250810925O} series on the Helmholtz \emph{Blablador} platform\footnote{\url{https://strube1.pages.jsc.fz-juelich.de/2024-02-talk-lips-blablador/}}\footnote{\url{https://sdlaml.pages.jsc.fz-juelich.de/ai/guides/blablador_api_access/}}, which exposes an OpenAI-compatible API and allows researchers to run models locally under institutional control. This makes the agent agnostic to the hosting model—cloud or on-prem—and enables systematic comparisons between proprietary and open-weight backends in our benchmark harness.

Future work includes expanding the RAG corpus (e.g.\ CTAO simulations) for end-to-end sensitivity studies, exploring multi-agent collaboration (planner, critic, optimiser) to improve robustness and convergence, and hardening open-weight deployments (quantization, schedulers, and multi-GPU inference) on Blablador. We plan to continue releasing the package and benchmarks to support transparent, reproducible research across both closed and open ecosystems.

\section*{Acknowledgements}
This work makes use of data from the H.E.S.S. DL3 public test data release~1, analyzed with the Gammapy framework. We thank Alexandre Strube for his assistance with the Blablador platform. The proceedings text is an experimental attempt at building an agent for automatic prose generation based on research results.

\bibliographystyle{JHEP}
\bibliography{references}

\end{document}